\documentclass{nature}

\usepackage{graphicx}
\usepackage[font=small]{caption}
\usepackage{xcolor}
\usepackage{amsmath} 
\usepackage{hyperref}
\hypersetup{
    colorlinks,
    linkcolor={blue!80!black},
    citecolor={blue!50!black},
    urlcolor={blue!80!black}
}

\title{Floquet-Bloch Valleytronics}

\author
{Sotirios Fragkos$^{1}$, Baptiste Fabre$^{1}$, Olena Tkach$^{2,3}$, Stéphane Petit$^{1}$, Dominique Descamps$^{1}$, Gerd Schönhense$^{2}$, Yann Mairesse$^{1}$, Michael Schüler$^{4,5}$, and Samuel Beaulieu$^{1}$}

\begin{document}

\maketitle

\begin{affiliations}
 \item Universit\'e de Bordeaux - CNRS - CEA, CELIA, UMR5107, F33405 Talence, France.
 \item Johannes Gutenberg-Universität, Institut für Physik, D-55099 Mainz, Germany.
 \item Sumy State University, Rymskogo-Korsakova 2, 40007 Sumy, Ukraine.
 \item PSI Center for Scientific Computing, Theory and Data, 5232 Villigen PSI, Switzerland.
 \item Department of Physics, University of Fribourg, CH-1700 Fribourg, Switzerland.
\end{affiliations}

\begin{abstract}
Driving quantum materials out of equilibrium makes it possible to generate states of matter inaccessible through standard equilibrium tuning methods. Upon time-periodic coherent driving of electrons using electromagnetic fields, the emergence of Floquet-Bloch states enables the creation and control of exotic quantum phases. In transition metal dichalcogenides, broken inversion symmetry within each monolayer results in a non-zero Berry curvature at the K and K$^{\prime}$ valley extrema, giving rise to chiroptical selection rules that are fundamental to valleytronics. Here, we bridge the gap between these two concepts and introduce Floquet-Bloch valleytronics. Using time- and polarization-resolved extreme ultraviolet momentum microscopy combined with state-of-the-art ab initio theory, we demonstrate the formation of valley-polarized Floquet-Bloch states in 2H-WSe$_2$ upon below-bandgap coherent electron driving with circularly-polarized light pulses. We investigate quantum path interference between Floquet-Bloch and Volkov states, showing that this interferometric process depends on the valley pseudospin and light polarization-state. Conducting extreme ultraviolet photoemission circular dichroism in these nonequilibrium settings reveals the potential for controlling the orbital character of Floquet-engineered states. These findings link Floquet engineering and quantum geometric light-matter coupling in two-dimensional materials. They can serve as a guideline for reaching novel out-of-equilibrium phases of matter by dynamically breaking symmetries through coherent dressing of winding Bloch electrons with tailored light pulses.
\end{abstract}

\section{Main}

Controlling electrons' valley degrees of freedom holds potential as a resource for (quantum) information processing~\cite{Culcer12}. In inversion-symmetry-broken systems, e.g., transition metal dichalcogenide (TMDC) monolayers, the momentum-space texture of the Berry curvature gives rise to specific chiroptical selection rules~\cite{Yao08}. The resonant photoexcitation of electrons from the valence (VB) to the conduction band (CB) with circularly polarized light occurs selectively within a given valley~\cite{Mak12, Zeng12, Cao12, Schaibley2016, Tyulnev24}--  the primary concept behind valleytronics. Standard equilibrium tuning strategies (e.g. electric~\cite{Mak14}, magnetic~\cite{Aivazian15}, mechanical~\cite{Zhu13}, and twistronics~\cite{Scuri20}) of valley degrees of freedom have led to breakthroughs in TMDC-based valleytronic devices~\cite{Radisavljevic2011}. Recent theoretical proposals suggest that nonequilibrium band structure engineering in a valley-specific manner would offer radically different all-optical control knobs for disruptive valleytronic applications~\cite{Kundu16, Claassen16, Lu21, Zhan22, Zhan23, Li24, Cao24}.

Tailoring quantum materials using light is a burgeoning field allowing unprecedented opportunities for generating new states of matter~\cite{Basov17, Torre21, Bao2022}. One of the most promising routes for nonequilibrium control of materials is Floquet engineering, which relies on the coherent dressing of matter via photonic time-periodic perturbations~\cite{Oka09, Kitagawa11, Lindner11, Wang13, Sie15, Mahmood16, Oka19, Rudner20, Reutzel20, Shan21, McIver20,  Aeschlimann21,  Zhou23, Zhou23_2, Ito23, Kobayashi23, liu23, weitz24, Bao2024}. These perturbations can be tailored by shaping the driving field in amplitude, frequency, and polarization to produce a specific targeted dressed quantum state of matter~\cite{Claassen16, Hubener17, Lui18, Schuler22, Trevisan22, Strobel23, Zhou23, Zhou23_2}. For example, valley-specific Floquet-engineering using circularly polarized light is at the origin of valley-dependent optical Stark effect and Bloch-Siegert shift, which can be indirectly measured from shifts in the optical spectra of light-dressed TMDCs~\cite{Sie15, Sie17}. 

Time- and angle-resolved photoemission spectroscopy (trARPES) is the most direct technique to probe the eigenstates of driven materials by measuring the energy and momentum of electrons ejected from their surface. It is thus a powerful tool to investigate Floquet physics, as it enables the detection of light-induced energy gaps, band replica, band renormalizations, and Floquet-Volkov interferences~\cite{Park14, Mahmood16, choi24, merboldt24, Bao25}, as hallmarks for Floquet band creation~\cite{Wang13, Mahmood16, Ito23, Zhou23, Zhou23_2, choi24, merboldt24, Bao2024}. This technique also provides a path to discriminating between photoinduced band structure engineering and population effects, typically entangled in time-resolved transport experiments~\cite{Sato19, Murotani23, McIver20} and complicating the determination of microscopic origins of light-induced phenomena. Interferometric photoemission methods have also been shown to be well suited to probe light-induced phenomena such as control of spin polarization in multiphoton photoemission from Cu(001)
~\cite{Winkelmann08} and nonlinear photoelectric responses in noble metals~\cite{Reutzel20_2}, for example. 

With strategies for the unambiguous detection of Floquet-Bloch bands via photoemission now established, the next important step is to move beyond mere observation to explore novel symmetry-engineered Floquet-Bloch phases of matter. However, a direct experimental demonstration of valley-specific Floquet-Bloch engineering remains an open challenge.

\begin{figure}
\centering\includegraphics[width=\textwidth]{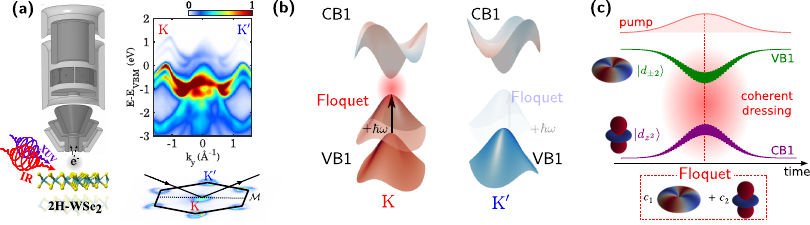}
\caption{\textbf{Experimental Scheme and Concept of Floquet Valleytronics}. \textbf{(a)} A polarization-tunable infrared pump (1.2~eV, 135~fs, 5.7 mJ/cm$^2$) and a polarization-tunable XUV (21.6~eV) probe pulses are focused on 2H-WSe$_2$, in the interaction chamber of a time-of-flight momentum microscope, at an incidence angle of 65$^{\circ}$ and with the light incidence plane along the crystal mirror plane $\mathcal{M}$ ($\Gamma$-M direction). The energy-momentum cut along K-$\Gamma$-K$^{\prime}$, recorded at pump-probe overlap and integrated over all pump IR quarter-wave plate (QWP) angles, shows the emergence of +$\hbar\omega$ sideband. The modulation of the IR pump polarization enables the investigation of valley-polarized Floquet states. In contrast, the modulation of the XUV polarization provides access to circular dichroism (CD-ARPES) from light-driven states. \textbf{(b)} Schematic of the Floquet-engineered electronic structure of 2H-WSe$_2$ around K and K$^{\prime}$ valleys. Upon time-periodic driving using IR circularly-polarized pulse, valley-polarized Floquet states are created. For simplicity, only the topmost valence band (VB1), associated +$\hbar\omega$ sideband, and first conduction band (CB1) are shown. Here, Floquet-Bloch Valleytronics refers to the light-induced asymmetric population of Floquet-Bloch bands with distinct valley pseudospins, specifically K and K$^\prime$, driven by different light helicities. \textbf{(c)} Schematic of the population dynamics of VB1 and CB1 induced by sub-gap coherent dressing and orbital hybridization associated with the formation of Floquet sideband.}
\label{Fig1}
\end{figure}

In this article, we use time- and polarization-resolved extreme ultraviolet (XUV) momentum microscopy combined with time-dependent non-equilibrium Green's function (td-NEGF) calculations to unambiguously demonstrate the creation of strongly valley-polarized Floquet-Bloch states in transition metal dichalcogenide 2H-WSe$_2$, upon below-bandgap pumping using circularly-polarized light pulses. We show that the emergence of these Floquet bands leaves their imprint as a polarization- and valley-dependent modulation of photoelectron angular distributions, due to quantum path interference with Volkov states. In addition, we perform XUV circular dichroism (CD-ARPES) from these light-dressed states and reveal the modification of orbital characters of Floquet-engineered states. Our work bridges the gap between Floquet-engineering and valleytronics and establishes a general methodology towards nonequilibrium engineering of broken symmetry quantum phases of matter using tailored light. 

\subsection{Concept of Floquet-Bloch Valleytronics} 

Figure~\ref{Fig1} shows a schematic of the experimental setup and the concept of Floquet-Bloch valleytronics. A polarization-tunable infrared (IR, 1.2~eV, 135~fs, $\sim$ 5.7 mJ/cm$^2$) pump and polarization-tunable XUV (21.6~eV) femtosecond probe pulses are focused onto a 2H-WSe$_2$ sample at room temperature in the interaction chamber of a time-of-flight momentum microscope. The incidence angle on the sample is 65$^{\circ}$ and the light incidence plane is along the $\Gamma$-M crystal mirror plane (for more details, see Methods). In the bulk limit, semiconducting 2H-WSe$_2$ is inversion-symmetric. However, it is characterized by locally broken inversion symmetry within each layer, which combined with strong spin-orbit coupling leads to coupled layer, spin, orbital, and valley degrees of freedom~\cite{Zhang14}. The surface sensitivity of XUV photoemission enables probing hidden spin~\cite{Riley14, Razzoli17, Fanciulli23} and orbital~\cite{Beaulieu20-2, Schuler22} texture, i.e. those of the top-most layer of the bulk crystal. This material is thus an ideal candidate for investigating valley-selective Floquet engineering. Figure~\ref{Fig1}(b) shows a schematic of the circularly-polarized light-driven electronic structure of the topmost layer of 2H-WSe$_2$, featuring the creation of valley-polarized Floquet states' populations (Floquet-Bloch valleytronics). 

\subsection{Valley- and Polarization-Resolved Quantum Path Interferences} 

\begin{figure}
\centering\includegraphics[width=\textwidth]{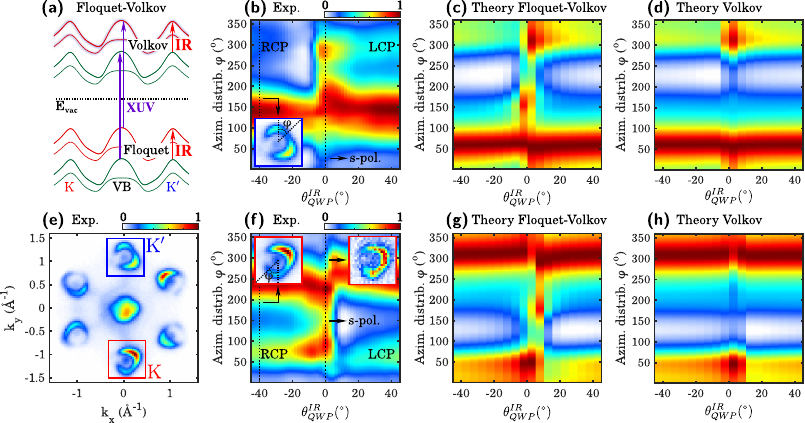}
\caption{\textbf{Valley- and Polarization-Resolved Quantum Path Interference Between Floquet-Bloch and Volkov States}. 
\textbf{(a)} Schematic of different coherent light-matter dressing effects, i.e. Floquet, Volkov, and quantum path interferences between Floquet and Volkov transitions. \textbf{(b)} and \textbf{(f)} Experimentally measured normalized azimuthal angular distribution of the photoemission intensity around K$^{\prime}$ and K valleys, respectively, of the +$\hbar\omega$ sideband (shown in \textbf{(e)}) as a function of the IR QWP angle ($\theta_{\mathrm{QWP}}^{IR}$), for $p-$polarized XUV probe, at the pump-probe temporal overlap. Insets represent the photoemission intensity around the K$^{\prime}$ and K valleys of the +$\hbar\omega$ sideband, for quasi-circular and $s-$polarized pumps (cut through black dashed lines). \textbf{(e)} Constant energy contour of the first-order sideband ($\mathrm{E-E_{VBM}}$ = 0.85~eV), integrated over all $\theta_{\mathrm{QWP}}^{IR}$. \textbf{(c)-(d)} and \textbf{(g)-(h)} Theoretical equivalent of \textbf{(b)} and \textbf{(f)}, including the contribution of both Floquet and Volkov transitions (\textbf{(c)} and \textbf{(g)}), and contribution from Volkov only (\textbf{(d)} and \textbf{(h)}).}
\label{Fig2}
\end{figure}

While Floquet states appear upon time-periodic dressing of materials' electronic band structure, Volkov states (also known as laser-assisted photoemission - LAPE) emerge from the coherent interaction between the pump field and outgoing photoelectron, an inherent process in trARPES. Both transitions end at the same final energy and in-plane momentum, leading to interferences (Fig.~\ref{Fig2}(a)). It was recently shown that quantum path interferences between Floquet and Volkov transitions manifest as pump-polarization-dependent modifications in the angular distribution of photoelectrons~\cite{Park14, Mahmood16, choi24, merboldt24}. This phenomenon provides an unambiguous method to identify the presence of Floquet bands. This measurement protocol lifts the challenge of measuring small energy gaps and band renormalizations as hallmarks for Floquet band formation. 

Following this method, we start by investigating the evolution of the light-induced sidebands in 2H-WSe$_2$ as a function of the IR pump polarization state. The laser ellipticity $\epsilon$ is modulated by controlling the quarter-wave plate angle ($\theta_{\mathrm{QWP}}^{IR}$), from circular-right ($\theta_{\mathrm{QWP}}^{IR}$ = -45$^{\circ}$, $\epsilon$=-1), to linear $s-$ ($\theta_{\mathrm{QWP}}^{IR}$ = 0$^{\circ}$, $\epsilon$=0), to circular-left ($\theta_{\mathrm{QWP}}^{IR}$ = +45$^{\circ}$, $\epsilon$=+1), where $\epsilon = \text{tan}(\theta_{\mathrm{QWP}}^{IR})$. We record the energy- and momentum-resolved photoemission intensity as a function of $\theta_{\mathrm{QWP}}^{IR}$, to obtain a four-dimensional dataset $\mathrm{I(E, k_x, k_y, \theta_{\mathrm{QWP}}^{IR}}$). Fig.~\ref{Fig1}(a) shows an energy-momentum cut along the $\mathrm{K-\Gamma-K^{\prime}}$ high-symmetry direction, integrated over all IR QWP angles. One notices the emergence of a first-order VB replica at an energy of +$\hbar\omega$ above the VB. At this relatively high pump fluence, multiphoton transitions also lead to a small but nonvanishing population of excited states in the conduction band (not visible here; see Extended Data Fig.~\ref{Sigma}). Moreover, in Extended Data Fig.~\ref{PPsp} and \ref{PPlr}, the emergence of the first-order valence band (VB) replica can be observed for different linear and circular pump polarizations, respectively, by taking the difference between the photoemission signals measured at the pump-probe overlap and before it.

In Fig.\ref{Fig2}(e), a constant energy contour (CEC) depicts the angular distribution of photoelectrons of the first-order sideband (+$\hbar\omega$), integrated over all QWP angles. The $\mathrm{k_x, k_y}$ distribution around each K/K$^{\prime}$ point of the +$\hbar\omega$ sideband is almost identical to the one from the VB (see Extended Data Fig.~\ref{CECs}). It is characterized by well-known croissant-shaped patterns, featuring a remarkable minimum in the photoemission yield, also called \textit{dark corridor}. It originates from interference between photoelectrons emitted from the W $d$-orbitals~\cite{Rostami19, Beaulieu20-2, Schuler22}. In Fig.~\ref{Fig2}(b) and (f), we track the evolution of the angular distribution around K$^{\prime}$/K valleys (red/blue boxes in Fig.~\ref{Fig2}(e)) of the +$\hbar\omega$ sideband, as a function of $\theta_{\mathrm{QWP}}^{IR}$. To do so, we plot the signal around K/K$^{\prime}$ in polar coordinates, normalized for each QWP angle. For both valleys, we notice a strong modification of the angular distribution around the $s-$polarization state configuration, where Floquet/Volkov contribution ratio is maximized (see Extended Data Figure~\ref{FloqVolk}). Indeed, one notices a stronger photoemission intensity emerging around 300$^{\circ}$ for K$^{\prime}$ (Fig.~\ref{Fig2}(b)) and around 65$^{\circ}$) for K valley (Fig.~\ref{Fig2}(f)). In addition, this new feature around different valley pseudospin exhibits opposite behavior when going away from $s-$polarization. For K$^{\prime}$ valley (Fig.~\ref{Fig2}(b)), it shows a sharp decay when going toward RCP IR driving pulses (negative $\theta_{\mathrm{QWP}}^{IR}$) and a smooth decay when going toward LCP IR driving pulses (positive $\theta_{\mathrm{QWP}}^{IR}$). This behavior is opposite for K valley (Fig.~\ref{Fig2}(f)): it shows a smooth decay when going toward RCP IR driving pulses (negative $\theta_{\mathrm{QWP}}^{IR}$) and a sharp decay when going toward LCP IR driving pulses (positive $\theta_{\mathrm{QWP}}^{IR}$). As reported in Bi$_2$Se$_3$~\cite{Mahmood16} and graphene~\cite{choi24,merboldt24}, this modification of the angular distribution is associated with quantum path interference between Floquet and Volkov transitions. 

To confirm the origin of these quantum path interferences, we performed trARPES simulations based on the time-dependent nonequilibrium Green's function (td-NEGF) approach. The light-matter interaction for both pump and probe pulses is treated from first principles, including photoemission matrix elements (see Methods). In Fig.~\ref{Fig2}(c)-(d) and (g)-(h), we present the theoretical evolution of the angular distribution around the K$^{\prime}$ and K valleys, showing both the combined Floquet and Volkov contributions, as well as the Volkov-only contribution. 
For $\theta_{\mathrm{QWP}}^{\mathrm{IR}} \neq 0^{\circ}$ (i.e., away from $s$-polarization), the Volkov contribution rapidly increases. As a result, Floquet-Volkov interferences are primarily observable near $s$-polarization, where Floquet and Volkov contributions to sideband amplitude are comparable. Away from $\theta_{\mathrm{QWP}}^{\mathrm{IR}} = 0^{\circ}$, the angular distributions from simulations including both Floquet and Volkov contributions and those including only Volkov transitions become essentially indistinguishable. 
Although simulations with only Volkov transitions reproduce most experimentally observed features, including Floquet transitions provide an even better match, especially for capturing subtle modulations near $s-$polarization. This includes the enhanced photoemission intensity observed around $\varphi = 150^{\circ}$--$200^{\circ}$ for an $s$-polarized driving field (see also Extended Data Fig.~\ref{AziTheo} for results including only Floquet transitions). These results unambiguously demonstrate the creation of Floquet states in 2H-WSe$_2$, their quantum path interference with Volkov states, as well as the possibility of controlling this interferometric process in a valley- and polarization-dependent manner. Indeed, in contrast to graphene, the K and K$^\prime$ valleys are not equivalent: the valence band maximum consists of a superposition of $d_{\pm 2}$ and $d_{z^2}$ orbitals, which are distinct at K and K$^\prime$, respectively. This unique orbital texture gives rise to the valley-dependent Volkov-Floquet interferences, which cannot be accounted for by simple geometric arguments. Indeed, as shown in previous work~\cite{Beaulieu20-2}, the orientation of the croissant-shaped photoemission angular distribution of the valence band (and here the sideband) is determined by the orbital pseudospin, or in other words, the valley-dependent complex superposition of $|d_{\pm 2}$ and $|d_{z^2}\rangle$ orbitals. The angular distribution presented in Fig.~\ref{Fig2} shows an interplay of orbital character of the valence band (and thus the orbital pseudospin) and the Floquet-Volkov interference close to $s$-polarization.

\subsection{Valley-Polarized Floquet-Bloch States} 

\begin{figure}
\centering\includegraphics[width=0.8\textwidth]{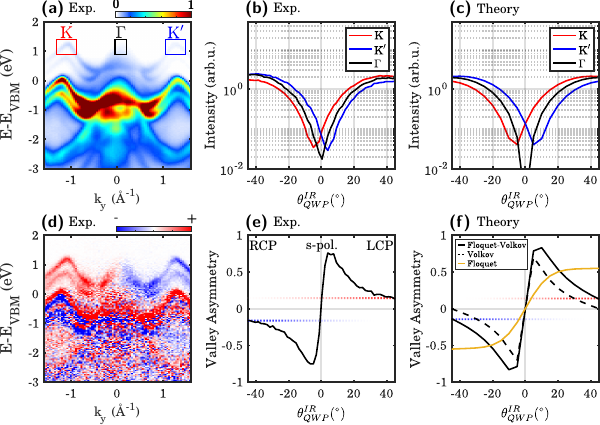}
\caption{\textbf{Valley-Polarized Floquet-Bloch States in 2H-WSe$_2$}. \textbf{(a)} Energy-momentum cut along K-$\Gamma$-K$^{\prime}$ direction, measured at the pump-probe temporal overlap and integrated for all IR quarter-wave plate angles ($\theta_{\mathrm{QWP}}^{IR}$). \textbf{(b)} Normalized photoemission intensity from the +$\hbar\omega$ sideband around $\Gamma$ (black line), K (red line) and K$^{\prime}$ (blue line) valleys, as a function of $\theta_{\mathrm{QWP}}^{IR}$. \textbf{(c)} Theoretical equivalent of \textbf{(b)}, including the contribution of both Floquet and Volkov transitions. \textbf{(d)} Differential energy-momentum cut along K-$\Gamma$-K$^{\prime}$ direction, measured at pump-probe overlap, obtained by subtracting the photoemission spectra measured using right- (RCP) and left-circularly (LCP) polarized IR driving pulses, highlights the emergence of valley-polarized +$\hbar\omega$ replica. \textbf{(e)} Polarization-resolved valley asymmetry of the +$\hbar\omega$ sideband, extracted through the normalized difference between photoemission intensity at K and K$^{\prime}$ shown in \textbf{(b)}. \textbf{(f)} Same as \textbf{(e)}, but using the calculated photoemission intensities, when including the contribution of Floquet only (solid yellow line), Volkov only (dashed black line)  and coherent sum of Floquet and Volkov (solid black line).}
\label{Fig3}
\end{figure}

After establishing the emergence of Floquet-Bloch states, we turn our attention to the valley-resolved evolution of their population as a function of the helicity of the driving field. Figure~\ref{Fig3}(b), shows the photoemission intensity from the +$\hbar\omega$ sideband around $\Gamma$, K, and K$^{\prime}$ valleys (black, red, and blue boxes, respectively, in Fig.~\ref{Fig3}(a)), as a function of $\theta_{\mathrm{QWP}}^{IR}$. The signal at $\Gamma$ is minimized when using $s-$polarization ($\theta_{\mathrm{QWP}}^{IR}$ = 0$^{\circ}$) and grows symmetrically when going towards opposite light helicity ($\theta_{\mathrm{QWP}}^{IR}$ = $\pm$45$^{\circ}$). This is the expected behavior for Volkov transitions, which are known to be proportional to the out-of-plane component of the laser electric field (Extended Data Fig.~\ref{FloqVolk} and Extended Data Fig.\ref{fig:volkov}). This IR polarization-dependent yield is strikingly different for K/K$^{\prime}$ valleys. Indeed, the +$\hbar\omega$ sideband intensity is stronger for right- (left-) IR circularly-polarized light-dressing for K (K$^{\prime}$) valley, and is minimized slightly before (after) reaching $s-$polarization. This behavior is further highlighted by plotting the valley asymmetry of the +$\hbar\omega$ sideband as a function of $\theta_{\mathrm{QWP}}^{IR}$ (Fig.~\ref{Fig3}(e)). This asymmetry exhibits a sign flip when going through $s-$polarization, and converges to $\sim\pm$15$\%$ when using a right- or left-circularly polarized IR dressing field. The valley-polarization of these engineered Floquet-Bloch states can also be visualized in the differential map calculated by subtracting photoemission spectra obtained using right- and left-handed IR field, in Fig.~\ref{Fig3}(d), leading to the opposite positive/negative (red/blue) differential photoemission intensities at K$^{\prime}$ and K. 

Our theoretical calculation (Fig.~\ref{Fig3}(c) and (f)) captures very well the polarization-dependent sideband intensity, as well as the associated valley asymmetry. Decomposing the polarization-dependent valley asymmetry into contributions from Floquet and Volkov states allows us to draw some conclusions about the origin of this signal. First, the strong asymmetry for small light ellipticities (near $s-$polarization) is dominated by Volkov contribution and is of purely geometric origin~\cite{choi24, merboldt24}. Second, the formation of Volkov states does not depend on the light helicity and thus does not yield any valley polarization when using circularly polarized light. Last, photonic dressing using circularly-polarized light leads to strongly valley-polarized Floquet-Bloch populations, i.e. $>50\%$ under our experimental conditions. This pronounced valley polarization is reminiscent of the selection rules for interband transitions, hinting at an interplay of VB and CB, which is also at the origin of the valley optical Stark~\cite{kim14, Sie15} and Bloch-Siegert effects~\cite{Sie17}. Indeed, the dressed-state picture reveals that the valley-dependent interband dipole transition matrix element governs the strength of valley-selective coherent Floquet dressing. Owing to the valley-selectivity of interband transitions under circularly polarized pumping, the coherent dressing strength—and consequently the Floquet sideband intensity—becomes valley dependent. This mechanism accounts for the pronounced Floquet valley asymmetry observed as a function of the QWP angle in Fig.~\ref{Fig3}(f). Capturing such Floquet-Bloch valley polarization thus requires accounting for both time-periodic photonic dressing and concomitant dynamics governed by selection rules. This is achieved through our td-NEGF approach, which extends beyond standard Floquet theory (see Extended Data Fig.~\ref{floqchirality}).

\subsection{XUV Photoemission Circular Dichroism from Light-Dressed States} 

\begin{figure}
\centering\includegraphics[width=0.8\textwidth]{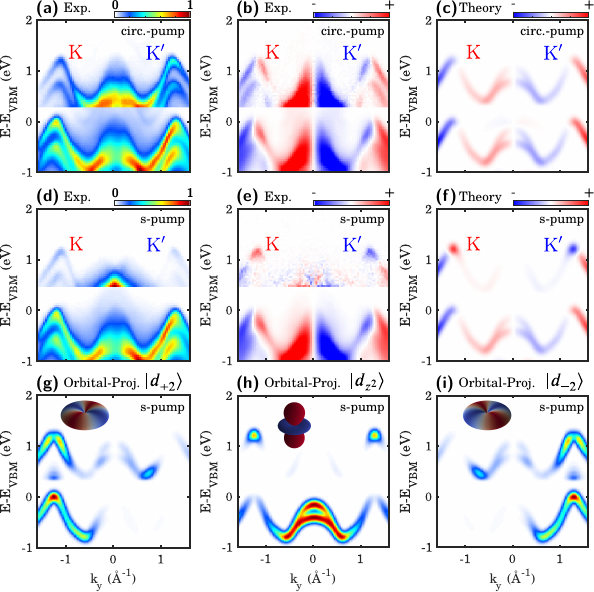}
\caption{\textbf{XUV Photoemission Circular Dichroism and Orbital Character of Light-Dressed States}. \textbf{(a)} Experimentally measured energy-momentum cut along K-$\Gamma$-K$^{\prime}$ direction, at pump-probe overlap, using right-circularly-polarized IR pump and integrated for all XUV quarter-wave plate angles, and \textbf{(b)} associated CD-ARPES. \textbf{(c)} Calculated CD-ARPES for circularly-polarized IR pump, including the contribution of both Floquet and Volkov transitions using theoretical approaches described in Methods. \textbf{(d)}-\textbf{f)} Same as \textbf{(a)}-\textbf{(c)}, but using $s-$polarized IR pump. \textbf{(g)}-\textbf{(i)} Orbital-resolved contributions to the trARPES signal along K-$\Gamma$-K$^{\prime}$ direction, using $s-$polarized IR pump. For each subpanel, the signal associated with the +$\hbar\omega$ sideband has been multiplied such that its absolute intensity matches the one from the VB.}
\label{Fig4}
\end{figure}

Now that we have demonstrated the creation of valley-polarized Floquet states, we performed advanced characterization of these light-engineered states using XUV circular dichroism (CD-ARPES). CD-ARPES is a powerful probe of the orbital angular momentum (OAM) and Berry curvature of Bloch states~\cite{Cho18, Schuler20, Ünzelmann2021, Yen2024}, even if experimental geometry, final state effects, and scattering can add complexity to the interpretation of the origin of the signals~\cite{Schonhense90, moser23, boban2024}. For 2H-WSe$_2$, the valence and conduction bands are expected to exhibit opposite circular dichroism, as their OAM are opposite. However, up to now, XUV CD-ARPES has been restricted to the investigation of ground-state equilibrium properties of solids. Here, we report the first XUV CD-ARPES from nonequilibrium states and investigate the modification of orbital texture in Floquet-engineered states of matter. To this end, we measure the modulation of photoemission intensity when switching the XUV light helicity, using our recently built all-reflective XUV quarter-wave plate~\cite{Comby22}, for two IR pump-polarization configurations ($s-$, and circular polarization). When driving 2H-WSe$_2$ using circularly polarized IR (Fig.~\ref{Fig4}(a)), the +$\hbar\omega$ sideband mainly originates from Volkov transitions (Extended Data Fig.~\ref{FloqVolk}). In that configuration, the CD-ARPES of the +$\hbar\omega$ sidebands is identical to the dichroism from the VB, i.e. is mainly characterized by a sign reversal exactly at the top of VB at K/K$^{\prime}$ (Fig.~\ref{Fig4}(b)). The situation is drastically different when using $s-$polarized IR dressing (Fig.~\ref{Fig4}(d)), which is the configuration maximizing the Floquet/Volkov amplitude ratio (Extended Data Fig.~\ref{FloqVolk}). In this case, CD-ARPES of the +$\hbar\omega$ sideband is strongly modified. Indeed, it is positive (negative) for K (K$^{\prime}$) valley at the VB maximum, and exhibits a sign reversal when going towards higher momenta and lower energy (Fig.~\ref{Fig4}(e)). This behavior is also reproduced in theoretical calculations (Fig.~\ref{Fig4}(c) and (f)). 

The CD-ARPES modification of the +$\hbar\omega$ sideband as compared to the VB is a clear indication that the orbital character has changed. To determine the origin of this behavior, we computed the orbital-resolved contributions to the trARPES signal (Fig.~\ref{Fig4}(g)-(i)). The VB and the largest part of the sideband are dominated by $|d_{\pm 2}\rangle$ orbitals at K/K$^\prime$. However, the orbital weight of $|d_{\pm 2}\rangle$ shows a dip at the top of the sideband, where a strong $|d_{z^2}\rangle$ contribution manifests. Since $|d_{z^2}\rangle$ is the orbital character of the CB minimum at K/K$^\prime$, we conclude that there is strong momentum-dependent hybridization of VB and CB in the +$\hbar\omega$ sideband. Within standard Floquet theory, weak-field, below-band-gap dressing does not account for the observed momentum-dependent hybridization (Extended Data Fig.~\ref{floqorb}), in contrast to our td-NEGF calculations. This discrepancy indicates that the hybridization effect arises from real-time population dynamics. To observe Floquet hybridization effects within the conventional Floquet spectral function, one must approach near-resonant, strong-field pumping conditions (Extended Data Fig.~\ref{floqorb}). 

In the weak-driving limit, population dynamics still occur while the pump pulse is present (see Fig.~\ref{Fig1}(c)). The induced CB population is predominantly virtual and decays to nearly zero once the pump pulse has ended. In other words, the VB and CB become dressed by the pump photons during the pulse. The hybridization effect can be understood from parity arguments: the virtual state dressed by one photon must have a parity opposite to the VB's ($P_\mathrm{VB}=1$). This is only possible if the virtual state has the parity of the CB ($P_\mathrm{CB}=-1$). This argument can be formalized by time-dependent perturbation theory (see Methods). Within this framework, this hybridization effect does require resonant driving. Our td-NEGF calculations naturally include the parity-governed population dynamics as well as non-perturbative Floquet effects (which are responsible for the sideband intensity away from K/K$^\prime$). Importantly, such hybridization effects in the off-resonant regime are signatures of coherent light-matter dressing and dynamics governed by selection rules. A Floquet steady state that forms in the presence of dissipation would not exhibit this type of significant band-character mixing (see Extended Data Fig.~\ref{floqorb}). On a more general level, the non-resonant band hybridization is driven by the momentum dependence of the VB-CB dipole transition matrix element $\mathbf{D}(\mathbf{k})$. The modern quantum geometric perspective on $\mathbf{D}(\mathbf{k})$ establishes a link between dipole transitions and the Berry connection~\cite{Ahn2022}, thus to topological properties. Indeed, the band hybridization in the +$\hbar\omega$ sideband can also be related to the quantum geometry of the Bloch states. 

\subsection{Conclusions}

Our time- and polarization-resolved momentum microscopy results complemented with time-dependent non-equilibrium Green's function calculations on light-driven 2H-WSe$_2$ unambiguously reveal the formation of Floquet-Bloch states through their polarization- and valley-dependent quantum path interference with Volkov states, their valley-polarized populations, and their orbital-texture modification. These results shed light on the interplay between Floquet-Bloch engineering and valley-selective populations emerging from selection rules when using circularly-polarized drivers that break time-reversal symmetry. These valley-polarized Floquet-Bloch states' populations offer great possibilities to control anomalous Hall effects in nonequilibrium settings using tailored nonresonant pulses~\cite{Cao24}. Moreover, the XUV CD-ARPES measurements in nonequilibrium settings reveal the hybridization of valence and conduction bands' orbitals underlying the formation of Floquet-Bloch states in the below-bandgap pumping regime, which was identified to be at the origin of valley-selective optical Stark and Bloch-Siegert shifts~\cite{kim14, Sie15, Sie17}. These results bridge the gap between Floquet engineering and quantum geometric light-matter coupling in TMDCs, greatly advancing the research frontier of nonequilibrium engineering of quantum materials.

\clearpage

\begin{methods}
\subsection{Time- and polarization-resolved momentum microscopy}
The experimental setup is articulated around a home-build polarization-tunable ultrafast XUV beamline~\cite{Comby22} coupled with a Momentum Microscope apparatus~\cite{Medjanik17}. The beamline is driven by a commercial high-repetition-rate Yb fiber laser (1030 nm, 135 fs (FWHM), 50 W, here operated at 166 kHz, Amplitude Laser Group). In the probe arm, we frequency-double a fraction of the beam in a BBO crystal and drive high-order harmonics by tightly focusing 515 nm pulses (5 W) in a thin and dense argon gas jet, in an annular beam geometry. We spatially separate the strong 515 nm driver and the XUV beamlet using spatial filtering with pinholes. We spectrally select the 9th harmonic (21.6 eV) by using a combination of reflections on SiC/Mg multilayer XUV mirrors (NTTAT) and transmission through a 200 nm thick Sn metallic filter. We control the XUV polarization states (linear $s-$ or $p-$, elliptical or quasi-circular) using a fixed all-reflective XUV quarter-wave plate made of four SiC mirrors (NTTAT) under 78$^{\circ}$ angle of incidence. In the pump arm, we use a small fraction of the fundamental laser pulse (1030 nm, 135 fs (FWHM), 115 mW, 0.69~$\mu$J/pulse). The IR pump and XUV probe pulses are recombined collinearly using a drilled mirror and focused onto the sample. The bulk $\mathrm{2H}$-$\mathrm{WSe_2}$ samples (HQ Graphene) were cleaved and introduced in a motorized 6-axis hexapod for sample alignment in the main chamber (base pressure of 2$\times$10$^{-10}$ mbar). The photoemission data are acquired using a custom time-of-flight momentum microscope equipped with a new front lens capable of different operating modes (GST mbH)~\cite{tkach2024multimode}. This detector allows for simultaneous detection of the full surface Brillouin zone, over an extended binding energy range, without the need to rearrange the sample geometry \cite{Medjanik17}. The momentum microscope settings used in the experiments allow for a momentum resolution of \(0.01~\text{\AA}^{-1}\), which has been previously determined using the Shockley surface state of Au(111).
 The temporal resolution of our setup is \(141~\text{fs}\), determined by a Gaussian fit of the cross-correlation signal (photoemission from the pump-induced sideband) between the pump and probe pulses (see Extended Data Fig.~\ref{Sigma}). At room temperature, where all measurements presented in this manuscript were performed, the total energy resolution is \(113~\text{meV}\). This total resolution, \(\Delta E_{\text{tot}}\), is determined by the convolution of the Fermi-Dirac distribution at a given temperature \(T\), the bandwidth of the photoionizing radiation (\(\Delta E_{h\nu}\)), and the resolution of the detector (\(\Delta E_{\text{ToF}}\)). The thermal width of the Fermi-Dirac distribution is given by \(\Delta E_{\text{therm}} \approx 4kT\), which is approximately \(100~\text{meV}\) at room temperature. Cooling the sample down to \(16~\text{K}\) improves the total energy resolution to \(44~\text{meV}\). For the data post-processing, we use an open-source workflow \cite{Xian20, Xian19_2} to efficiently convert the raw single-event-based datasets into binned calibrated data hypervolumes of the desired dimension, including axes calibration. 

\subsection{Beam size and fluence determination}
The Photoemission Electron Microscopy (PEEM) mode of our momentum microscope enables the acquisition of real-space images of photoelectrons emitted from the surface of our sample. We employed this mode to characterize the spatial profiles of both the pump and probe beams. Because photoemission induced by the XUV probe beam involves a single-photon process, the PEEM image directly represents the spatial footprint of the XUV beam. From these measurements, we determined the full-width at half-maximum (FWHM) of the XUV beam's footprint to be approximately 45~$\mu$m in the horizontal direction (in the plane of incidence) and 33~$\mu$m in the vertical direction (perpendicular to the plane of incidence). The larger horizontal footprint is a result of the beam’s 65$^\circ$ angle of incidence on the crystal surface. In contrast, photoemission driven by the IR pump beam occurs via a nonlinear multiphoton process. Consequently, accurately determining the IR beam's spatial footprint requires consideration of the process’s effective nonlinearity. To quantify this nonlinearity, we measured the total photoemission yield as a function of the pump pulse power. The IR pump beam footprint was then derived by correcting the measured multiphoton PEEM footprint by the square root of the effective nonlinearity. Using this approach, we calculated the FWHM footprint of the IR pump beam to be approximately 138~$\mu$m in the horizontal direction and 69~$\mu$m in the vertical direction. This detailed beam size characterization allowed us to estimate the \textit{in situ} fluence of the IR pump beam to be approximately 5.7~mJ/cm$^2$, corresponding to an intensity of 4$\times$10$^{10}$~W/cm$^2$. The same incident fluence was maintained for all measurements presented in this manuscript.

\subsection{Extraction of the XUV CD-ARPES from nonequilibrium states}
To control the polarization state of the femtosecond XUV pulses, we rotate a visible (515 nm) zero-order half-wave plate in the driving beam arm before high-order harmonic generation. Since the linear polarization axis angle of the driving pulses is transferred to the high-order harmonics, this allows us to control the polarization axis angle of the XUV radiation impinging on the fixed all-reflective XUV quarter-wave plate. Upon a full rotation (360$^{\circ}$) of the visible (515 nm) half-wave plate, we go through 4 cycles of $p-$, quasi-LCP, $s-$, quasi-RCP, $p-$polarization state of the XUV radiation (maximum ellipticities of $\sim$90$\%$~\cite{Comby22}). To generate high-statistics XUV CD-ARPES from (weakly populated) nonequilibrium states, we sit at the pump-probe temporal overlap and continuously rotate the visible (515 nm) half-wave plate, while measuring the associated photoemission intensity on-the-fly, for more than 24 hours, at a count rate $\sim$4$\times$10$^5$ electrons/s. Each electron event is tagged with the instantaneous position of the visible (515 nm) half-wave plate. Binning the data using the procedure described above yields a four-dimensional dataset, i.e. $\mathrm{I(E_B, k_x, k_y, \theta_{HWP}^{IR})}$. We perform energy- and momentum-resolved Fourier analysis along $\theta_{\mathrm{QWP}}^{IR}$ axis. We extract the polarization-modulated signal oscillating at the helicity-switch frequency, corresponding to the CD-ARPES signal. This type of Fourier measurement scheme, analogous to lock-in detection, allows isolating signals that are modulated at the helicity-switch frequency (CD-ARPES), efficiently rejecting all other frequency components coming from, e.g., linear dichroism, experimental geometry, or artifacts (e.g., imperfection of the polarization state).

\subsection{Theoretical calculations}

We performed density-functional theory (DFT) calculations for a 2H-WSe$_2$ bilayer using the \textsc{Quantum Espresso} package\cite{giannozzi_quantum_2009}. DFT calculations were carried out using the Perdew–Burke–Ernzerhof (PBE) exchange-correlation functional and the pseudopotentials from the \textsc{PseudoDojo} project~\cite{van_setten_pseudodojo_2018}.
As the next step, we constructed a Wannier Hamiltonian (including all orbitals from the pseudopotentials) using projective Wannier functions. The light-matter coupling is then included through the velocity matrix elements $\mathbf{v}_{\alpha\alpha^\prime}(\mathbf{k})$, which yields the time-dependent Hamiltonian in bands basis:
\begin{equation}
	\label{eq:wann_velo}
	H_{\alpha\alpha^\prime}(\mathbf{k},t) = \varepsilon_\alpha(\mathbf{k}) \delta_{\alpha\alpha^\prime} - \mathbf{A}_\mathrm{pump}(t)\cdot \mathbf{v}_{\alpha\alpha^\prime}(\mathbf{k}) + \frac{1}{2}\mathbf{A}_\mathrm{pump}(t)^2 \ .
\end{equation}
Here, $\varepsilon_\alpha(\mathbf{k})$ are the band energies, and $\mathbf{A}_\mathrm{pump}(t)$ is the vector potential of the pump pulse inside the sample.

Dielectric effects in the material are taken into account using the Fresnel equations (similar to ref.~\cite{merboldt24}). We assume a sharp interface between vacuum and sample, where the incoming electric field $\mathbf{E}^{\mathrm{vac}}_\mathrm{pump}(t)$ is partially transmitted and reflected, respectively. The effective field within the bulk is modeled by the transmitted field $\mathbf{E}^{\mathrm{t}}_\mathrm{pump}(t)$ (which then defines $\mathbf{A}_\mathrm{pump}(t)$). As discussed in ref.~\cite{keunecke_electromagnetic_2020}, the laser-assisted photoemission (LAPE) process is mostly sensitive to the electric field close to the surface, which is given by a superposition of incoming $\mathbf{E}^{\mathrm{vac}}_\mathrm{pump}(t)$  and reflected field $\mathbf{E}^{\mathrm{r}}_\mathrm{pump}(t)$: $\mathbf{E}^\mathrm{eff}_\mathrm{pump}(t) = \mathbf{E}^{\mathrm{vac}}_\mathrm{pump}(t) + \mathbf{E}^{\mathrm{r}}_\mathrm{pump}(t)$.

With these ingredients, we constructed the Floquet Hamiltonian 
\begin{equation}
    \label{eq:floq_ham}
    \mathcal{H}_{\alpha n, \alpha^\prime n^\prime}(\mathbf{k}) = \frac{1}{T_\mathrm{pump}} \int^{T_\mathrm{pump}}_0 dt e^{i (n - n^\prime) \omega_\mathrm{pump} t } H_{\alpha\alpha^\prime}(\mathbf{k},t) -  n \omega_\mathrm{pump} \delta_{n n^\prime} \ ,
\end{equation}
where $\omega_\mathrm{pump}$ is the pump frequency and $T_\mathrm{pump} = 2\pi / \omega_\mathrm{pump}$. Diagonalizing the Floquet Hamiltonian yields the quasi-energy bands $\tilde{\varepsilon}_\lambda(\mathbf{k})$ and the associated Floquet eigenmodes $|\phi_{\mathbf{k}\lambda}\rangle$. To visualize the orbital character of the Floquet states, it is convenient to introduce the Floquet spectra function $\hat{A}(\mathbf{k}, \omega) = \sum_\lambda |\phi_{\mathbf{k}\lambda}\rangle \langle \phi_{\mathbf{k}\lambda} | g(\omega - \tilde{\varepsilon}_\lambda(\mathbf{k}))$, where $g(\omega)$ is a Gaussian function. Plotting the trace $\mathrm{Tr}[\hat{A}(\mathbf{k},\omega)]$ yields all quasi-energy bands in a certain energy window (see Extended Data Fig.~\ref{floqchirality}); taking $A_j(\mathbf{k},\omega) = \langle \varphi_j | \hat{A}(\mathbf{k},\omega) | \varphi_j\rangle$ allows us to define the orbital-resolved Floquet spectral function with respect to specific orbitals $|\varphi_j\rangle$ (Extended Data Fig.~\ref{floqorb}).

We computed the two-time Green's function $G^<_{\alpha \alpha^\prime}(\mathbf{k}, t, t')$ as in ref.~\cite{merboldt24}.
Following ref.~\cite{schuler_theory_2021-1} we then computed the theoretical trARPES signal from
\begin{equation}
    \label{eq:green_trarpes}
    I(\mathbf{k}, E) = \mathrm{Im} \sum_{\alpha \alpha^\prime} \int^\infty_0 dt \int^t_0 dt' \, s(t) s(t') e^{-i\Phi(\mathbf{k}, E, t, t')} M_{\alpha^\prime}(\mathbf{k},E) G^<_{\alpha^\prime \alpha}(\mathbf{k}, t', t) M^*_{\alpha}(\mathbf{k}, E) \ .
\end{equation}
Here, $\Phi(\mathbf{k}, E, t, t')$ is a phase factor that depends on the photon energy of the probe pulse and the pump vector potential dressing the photoelectrons (computed from $\dot{\mathbf{A}}^\mathrm{eff}_\mathrm{pump}(t) = - \mathbf{E}^\mathrm{eff}_\mathrm{pump}(t)$ ). The envelope of the probe pulse is described by $s(t)$, for which we chose a Gaussian pulse with the same center as the pump pulse. The FWHM for both pulses is chosen as in the experiments. Finally, the photoemission matrix elements $M_{\alpha}(\mathbf{k},E)$ are computed as in ref.~\cite{yen_first-principle_2024}.

The different contributions to the trARPES signal can be disentangled by selectively replacing the pump field with zero in the light-matter coupling. Pure Floquet effects can be simulated by replacing $\mathbf{A}^\mathrm{eff}_\mathrm{pump}(t) \rightarrow 0$ in the phase factor $\Phi(\mathbf{k}, E, t, t')$ (which removes LAPE). Pure Volkov effects, on the other hand, can be studied by replacing the Green's function $G^<_{\alpha \alpha^\prime}(\mathbf{k}, t, t')$ by the field-free equilibrium version.

To gain a better understanding of the influence of the experimental geometry on the Volkov effects, we also employed a model inspired by Park~\cite{Park14}. We consider the Volkov contribution to the phase factor $\Phi(\mathbf{k}, E, t, t') = E (t - t') + \Phi_V(\mathbf{k}, E, t, t')$ with
\begin{align}
    \label{eq:volkov_phase}
    \Phi_V(\mathbf{k}, E, t, t') = \int^t_{t'}d \bar{t}\left[ -\mathbf{p}\cdot \mathbf{A}^\mathrm{eff}_\mathrm{pump}(\bar{t}) + \frac{1}{2}\mathbf{A}^\mathrm{eff}_\mathrm{pump}(\bar{t})^2 \right] \ .
\end{align}
The intensity of Volkov sidebands can be obtained by performing a Fourier analysis of the time-dependent phase factor $e^{-i \Phi_V(\mathbf{k}, E, t, 0)}$. Assuming the diamagnetic term is small and a time-periodic pump, this procedure yields the sideband intensity
\begin{align}
    \label{eq:bessel}
    I^V_n(\mathbf{k}, E) \propto J_n(\alpha(\mathbf{k}, E))^2 \ , \ \alpha(\mathbf{k}, E) = \frac{E^\mathrm{vac}_0}{\omega_\mathrm{pump}} \sqrt{(\hat{\epsilon}'\cdot \mathbf{p})^2 + (\hat{\epsilon}''\cdot \mathbf{p})^2 } \ .
\end{align}
Here, $\mathbf{p} = \mathbf{k} + p_\perp \hat{z}$ is the three-dimensional momentum of the photoelectron with $E = \mathbf{p}^2/2$, while $\hat{\epsilon} = \hat{\epsilon}' + i \hat{\epsilon}''$ is the complex polarization vector determined by the Fresnel equations. The predictions of Eq.~\eqref{eq:bessel} are consistent with the Volkov-only simulations given by Eq. 3 of the method section. In particular, for our experimental geometry and photon energy, Eq.~\eqref{eq:bessel} explains why the Volkov contribution increases with $|\theta^{IR}_\mathrm{QWP}|$: $\hat{\epsilon}$ acquires an out-of-plane component, increasing $\alpha(\mathbf{k},E)$. Similarly, at $\theta^{IR}_\mathrm{QWP}=0^{\circ}$ ($s$-polarization) and $\theta^{IR}_\mathrm{QWP}=\pm45^{\circ}$ (circular polarizations), $\hat{\epsilon}\cdot\mathbf{p}$ and thus, the Volkov signal is identical for K and K$^\prime$.

To illustrate the contributions of the individual orbitals in Fig.~\ref{Fig4}, we replaced the photoemission matrix elements $M_\alpha(\mathbf{k},E) \rightarrow U_{j\alpha}(\mathbf{k})$, where $U_{j\alpha}(\mathbf{k})$ is the projection of eigenvector of Wannier Hamiltonian onto a chosen orbital $j$. LAPE effects were switched off for this calculation.

The imprint of the interband selection rules on the $\hbar \omega_\mathrm{pump}$ sideband can also be understood from time-dependent perturbation theory. To this end we write the time-dependent wave-function as $|\phi_{\mathbf{k}}(t)\rangle = \sum_\alpha C_{\mathbf{k}\alpha}(t) | \psi_{\mathbf{k}\alpha}\rangle + \int dE\, C_{\mathbf{k},E}(t) |\chi_{\mathbf{k},E}\rangle$, where $| \psi_{\mathbf{k}\alpha}\rangle $ are the Bloch states of the relevant bands, and $|\chi_{\mathbf{k},E}\rangle$ are the photoelectron states. The light-matter coupling is introduced between VBs and CBs as $H^\mathrm{dip}_{\alpha \beta}(\mathbf{k},t) = - \mathbf{E}_\mathrm{pump}(t)\cdot \mathbf{D}_{\alpha \beta}(\mathbf{k})$; the interband dipole matrix element $\mathbf{D}_{\alpha \beta}(\mathbf{k})$ is identical to the Berry connection $\mathbf{A}_{\alpha\beta}(\mathbf{k})$. The probe pulse is included by $H^\mathrm{PES}_{E,\alpha}(\mathbf{k},t) = - E_\mathrm{probe}(t) M_\alpha(\mathbf{k},E)$. The photoemission intensity is then obtained from $ I(\mathbf{k},E) \propto |C_{\mathbf{k}\alpha}(t)|^2$ for $t\rightarrow \infty$. Employing second-order perturbation theory including the combined effect of $H^\mathrm{dip}_{\alpha \beta}(\mathbf{k},t)$ and $H^\mathrm{PES}_{E,\alpha}(\mathbf{k},t)$ yields an expression similar to Fermi's Golden rule:
\begin{equation}
    \label{eq:pt_intensity}
    I(\mathbf{k},E) \propto g(\varepsilon_\alpha(\mathbf{k}) + \omega_\mathrm{pump} - E) \left| T_\alpha(\mathbf{k},E)\right|^2 \ ,
\end{equation}
where $g(\omega)$ is the squared Fourier transform of the envelope $s(t)$. The effective matrix element is defined by
\begin{equation}
    \label{eq:pt_amplitude}
    T_\alpha(\mathbf{k},E) = \sum_\beta \frac{M_\beta(\mathbf{k}, E)\mathbf{e}_\mathrm{pump} \cdot \mathbf{A}_{\beta\alpha}(\mathbf{k})}{\varepsilon_\beta(\mathbf{k}) - \varepsilon_\alpha(\mathbf{k}) - \omega_\mathrm{pump} } \ .
\end{equation}
Here, we have assumed an infinitely long pump pulse with polarization $\mathbf{e}_\mathrm{pump} $. Eqs.~(\ref{eq:pt_intensity})--(\ref{eq:pt_amplitude}) demonstrate that for overlapping pump and probe pulse: (1) photoemission intensity is observed at $\varepsilon_\alpha(\mathbf{k}) + \omega_\mathrm{pump}$, i.\,e. a $\hbar\omega_\mathrm{pump}$ sideband of the VB, and (2) the intensity is governed by the interband Berry connection $ \mathbf{A}_{\beta\alpha}(\mathbf{k})$ with CBs $\beta$ and thus the valley selection rules. The denominator in Eq.~(\ref{eq:pt_amplitude}) renders this effect non-resonant, which is consistent with the experimental observations. We have also confirmed the absence of resonances by varying the pump photon energy $\hbar\omega_\mathrm{pump}$ in the full td-NEGF simulations (Extended Data Fig.~\ref{photoncdad}).

\end{methods}

\clearpage

\begin{addendum}
\item [Acknowledgement] We thank Nikita Fedorov, Romain Delos, Pierre Hericourt, Rodrigue Bouillaud, Laurent Merzeau, and Frank Blais for technical assistance. M.S. acknowledges support from SNSF Ambizione Grant No. PZ00P2-193527. This work was supported by state funding from the ANR under the France 2030 program, with reference ANR-23-EXLU-0004, PEPR LUMA TORNADO. We acknowledge the financial support of the IdEx University of Bordeaux/Grand Research Program "GPR LIGHT". We acknowledge support from ERC Starting Grant ERC-2022-STG No.101076639, Quantum Matter Bordeaux, AAP CNRS Tremplin and AAP SMR from Université de Bordeaux. Funded by the European Union. Views and opinions expressed are however those of the author(s) only and do not necessarily reflect those of the European Union. Neither the European Union nor the granting authority can be held responsible for them. 
 
 \item[Author contributions] 
S.B. and M.S. conceived the idea. S.B., S.F., and Y.M. performed the experiments. S.B. analyzed the experimental data. B.F. optimized the multidimensional binning code. D.D. and S.P. participated in maintaining the laser system. G.S. and O.T. designed, implemented, and optimized the momentum microscope. M.S. developed the theory. S.B. and M.S. wrote the manuscript with input from other authors.

 \item[Competing Interests] The authors declare that they have no
competing financial interests.

 \item[Correspondence] Correspondence and requests for materials
should be addressed to S.B. (samuel.beaulieu@u-bordeaux.fr) or M.S. (michael.schueler@psi.ch).

 \item[Data availability] The data that support the findings of this article are openly available on Zenodo 
 \url{https://doi.org/10.5281/zenodo.15221633}

 \end{addendum}

\clearpage

\setcounter{figure}{0} 
\renewcommand{\figurename}{Extended Data Fig.}

\begin{figure}
\centering\includegraphics[width=0.8\textwidth]{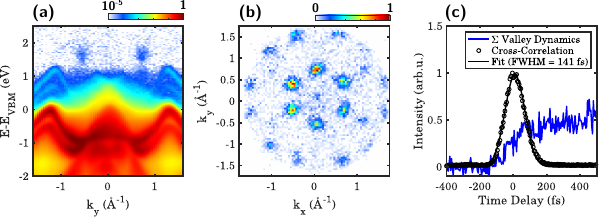}
\caption{\textbf{Floquet-Volkov states, two-photon excited states mapping and experimental cros$s-$correlation}. Using an $s-$polarized driving IR pump pulse (1.2~eV) and a $p-$polarized XUV probe, we map the concomitant emergence of Floquet-Volkov bands at K/K$^{\prime}$ valleys and excited-state populations in the $\Sigma$ valleys. \textbf{(a)} Energy-momentum cut along the K-$\Gamma$-K$^{\prime}$ high-symmetry direction, integrated over all pump-probe delays, shown on a false-color logarithmic scale. This cut reveals the appearance of first-order sidebands at K/K$^{\prime}$ valleys and excited-state populations in the $\Sigma$ valleys. Since we are operating in a below-band-gap pumping regime, the observation of excited-state populations in $\Sigma$ valleys highlights the significance of two-photon processes under these strong pumping conditions ($\sim$ 5.7 mJ/cm$^2$). \textbf{(b)} A constant energy contour for $\mathrm{E-E_{VBM}}$ above 1.3~eV and positive pump-probe delays illustrates the momentum distribution of long-lived excited states in the $\Sigma$ valleys, populated via a two-photon absorption process. \textbf{(c)} Time-resolved photoemission signal from the excited states in the $\Sigma$ valleys (blue line) are shown alongside the pump-probe cros$s-$correlation curve (obtained using $p-$polarized IR pump) with associated Gaussian fit. From this fit, the experimental temporal resolution is determined to be 141~fs full-width at half-maximum (FWHM).}
\label{Sigma}
\end{figure}

\begin{figure}
\centering\includegraphics[width=0.8\textwidth]{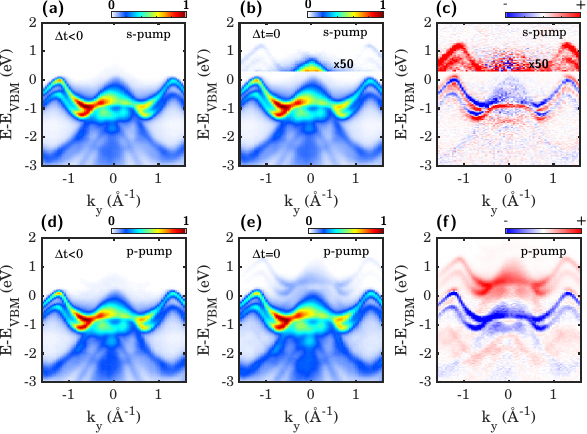}
\caption{\textbf{Dressed electronic structure for $s-$ and $p-$polarized driving pump pulses}. \textbf{(a), (b)} and \textbf{(c)} Energy-momentum cuts along K-$\Gamma$-K$^{\prime}$ high symmetry direction using $s-$polarized IR driving pump and $p-$polarized XUV probe, \textbf{(a)} before the temporal overlap \textbf{(b)} at the temporal overlap, and \textbf{(c)} differential map between \textbf{(a)} and \textbf{(b)}. \textbf{(d), (e)} and \textbf{(f)} Same as \textbf{(a), (b)} and \textbf{(c)}, but for $p-$polarized IR driving pulses. For $p-$polarized pump, the overall shape of the replica is extremely similar to the shape of the valence band. At the temporal overlap, the dominant effect in the valence band seems to be depletion i.e. appearing as a negative (blue) signal in the differential map. This is consistent with the Volkov mechanism, where the outgoing electrons are dressed by the out-of-plane component of light at the surface of the sample, leading to a redistribution of photoemission intensity at +$\hbar\omega$ and $-\hbar\omega$ with respect to the ground state valence band. The situation is drastically different for $s-$polarized driving pump pulses. The first-order sideband exhibits a different energy-momentum-dependent spectral weight than the valence band. Indeed, while the first-order sideband around K/K$^{\prime}$ points resembles the valence band dispersion rigidly shifted by +$\hbar\omega$, the situation seems more complex around $\Gamma$.}
\label{PPsp}
\end{figure}

\begin{figure}
\centering\includegraphics[width=\textwidth]{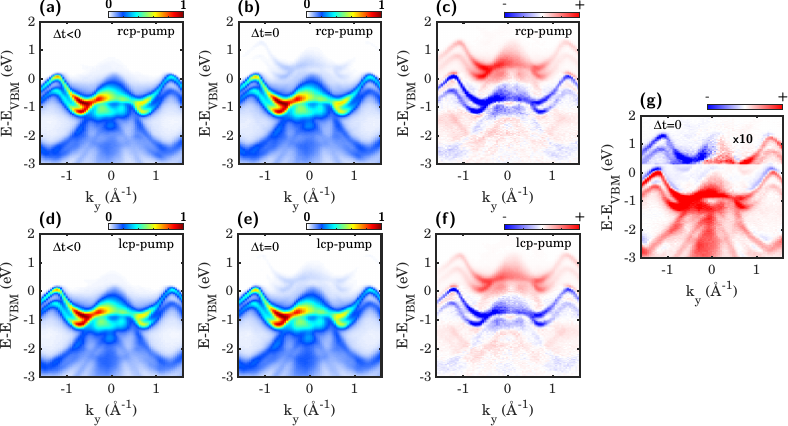} \caption{\textbf{Dressed electronic structure for right- and left-circularly polarized driving pump pulses}.\textbf{(a), (b)}, and \textbf{(c)} Energy-momentum cuts along the K-$\Gamma$-K$^{\prime}$ high-symmetry direction using right-circularly polarized (rcp-) driving pulses: \textbf{(a)} before temporal overlap, \textbf{(b)} at temporal overlap, and \textbf{(c)} the differential map between \textbf{(a)} and \textbf{(b)}. \textbf{(d), (e)}, and \textbf{(f)} Same as \textbf{(a), (b)}, and \textbf{(c)}, but for left-circularly polarized (lcp-) driving pulses. \textbf{(g)} Differential map between right- and left-circularly polarized pump pulses at the temporal overlap (i.e., the difference between \textbf{(b)} and \textbf{(e)}).
The out-of-equilibrium photoemission intensity using circularly polarized drivers is strikingly similar to that observed with $p-$polarized pump pulses. This similarity is not surprising, given that we have already established the greater efficiency of the Volkov mechanism with $p-$polarized light compared to the Floquet mechanism (see Extended Data Fig.~\ref{FloqVolk}). Consequently, as in the case of $p-$polarized drivers, the differential maps in \textbf{(c)} and \textbf{(f)} are characterized by replicas at +$\hbar\omega$ and $-\hbar\omega$, as well as a pronounced depletion of the valence band. \textbf{(g)} When taking the difference between right- and left-circularly polarized pump pulses at the temporal overlap, a positive (negative) differential signal is observed on the sideband at the K (K$^{\prime}$) valley. This phenomenon is reminiscent of the valley-polarized Floquet-Bloch bands generated by dressing this material with circularly-polarized light.}
\label{PPlr}
\end{figure}

\begin{figure}
\centering\includegraphics[width=0.8\textwidth]{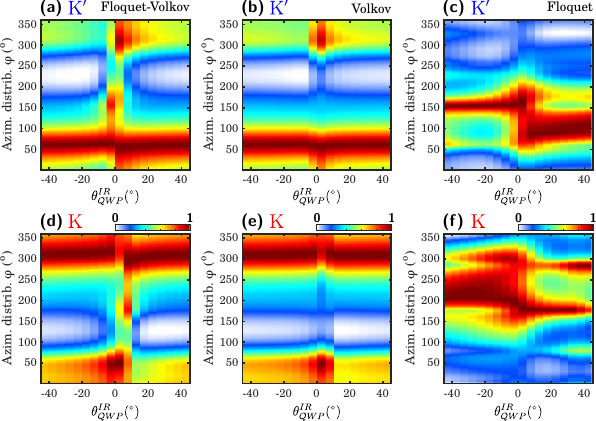}
\caption{\textbf{Floquet and Volkov contributions to valley- and polarization-resolved photoelectron angular distributions}. Calculated photoelectron angular distributions of the +$\hbar\omega$ sideband as a function of the IR QWP angle, for \textbf{(a),(d)} coherent sum of Floquet and Volkov transitions, \textbf{(b), (e)} Volkov transition only, and \textbf{(c), (f)} Floquet transition only, at K$^{\prime}$ and K valley, respectively. To generate these maps, we plot the signal around K/K$^{\prime}$ in polar coordinates, normalized for each $\theta_{\mathrm{QWP}}^{IR}$. For both valleys, when taking into account the coherent sum of Floquet and Volkov transition, we notice a strong modification of the angular distribution, including partial filling of the dark corridor around the $s-$polarization state configuration, where Floquet/Volkov contribution ratio is maximized. In addition, these new features around different valley pseudospin exhibit opposite behavior when going away from $s-$polarization: it shows a sharp (smooth) decay when going toward right- (left-) circularly-polarized IR driving pulses for K (K$^{\prime}$) valley. While the contribution of Volkov states to this polarization-dependent photoemission intensity is significant, the complex angular distribution of Floquet states has a capital role in reshaping this angular distribution. When coherently summing Floquet and Volkov transitions, the calculated signal matches the experimentally measured polarization-resolved angular distributions.}
\label{AziTheo}
\end{figure}

\begin{figure}
\centering\includegraphics[width=0.8\textwidth]{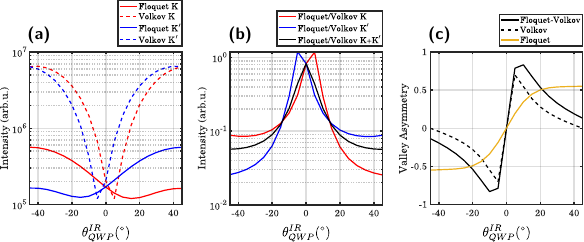}
\caption{\textbf{Calculated valley- and polarization-resolved Floquet and Volkov amplitudes}. \textbf{(a)} Calculated contribution of Floquet (solid lines) and Volkov transitions (dashed lines) to the photoemission intensity from the +$\hbar\omega$ sideband around K (red) and K$^{\prime}$ (blue) valleys, as a function of $\theta_{\mathrm{QWP}}^{IR}$. \textbf{(b)} Associated Floquet to Volkov ratio as a function of $\theta_{\mathrm{QWP}}^{IR}$, for K (red), K$^{\prime}$ (blue), as well as the sum of K (red) and K$^{\prime}$  valleys (black). \textbf{(c)} Calculated polarization-resolved valley asymmetry of the +$\hbar\omega$ replica, calculated through the normalized difference between photoemission intensity at K and K$^{\prime}$, when including the contribution of Floquet only (solid yellow line), Volkov only (dashed black line) and coherent sum of Floquet and Volkov (solid black line).}
\label{FloqVolk}
\end{figure}

\begin{figure}
\centering\includegraphics[width=0.8\textwidth]{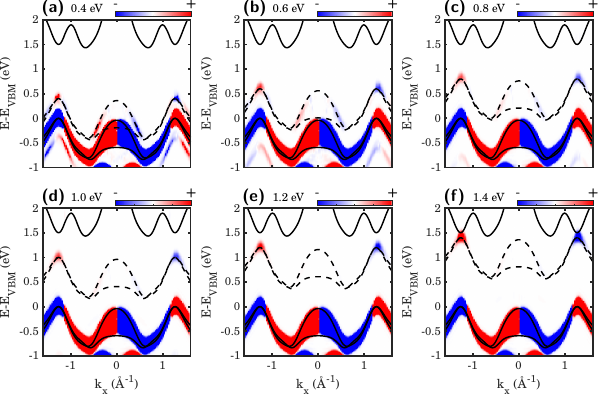}
\caption{\textbf{Calculated XUV CD-ARPES from light-induced Floquet states as a function of driving field detuning}. Using $s-$polarized pump and circularly polarized XUV probe field, we calculated the XUV CD-ARPES along K-$\Gamma$-K$^{\prime}$ high-symmetry direction as a function of pump photon energy to investigate the role of driving field detuning on the dichroic signal. The calculated static band structure (solid black line), as well as the top-most valence band rigidly shifted by the photon energy (dashed black line) are superimposed to the dichroic signal. For each photon energy, the colorbar is saturated at 0.1$\%$ of the maximum absolute dichroic signal from the valence band to enhance the visibility of Floquet states' dichroism. The pump photon energy is varied from \textbf{(a)} 0.4~eV, \textbf{(b)} 0.6~eV, \textbf{(c)} 0.8~eV, \textbf{(d)} 1.0~eV, \textbf{(e)} 1.2~eV, to \textbf{(f)} 1.4~eV. The dichroism texture near the band extrema of the sideband is stable against driving field detuning, suggesting that the orbital character of the Floquet-engineered bands is mainly dictated by the valence and conduction bands' orbital character and not the photon energy of the driver.}
\label{photoncdad}
\end{figure}

\begin{figure}
\centering\includegraphics[width=0.7\textwidth]{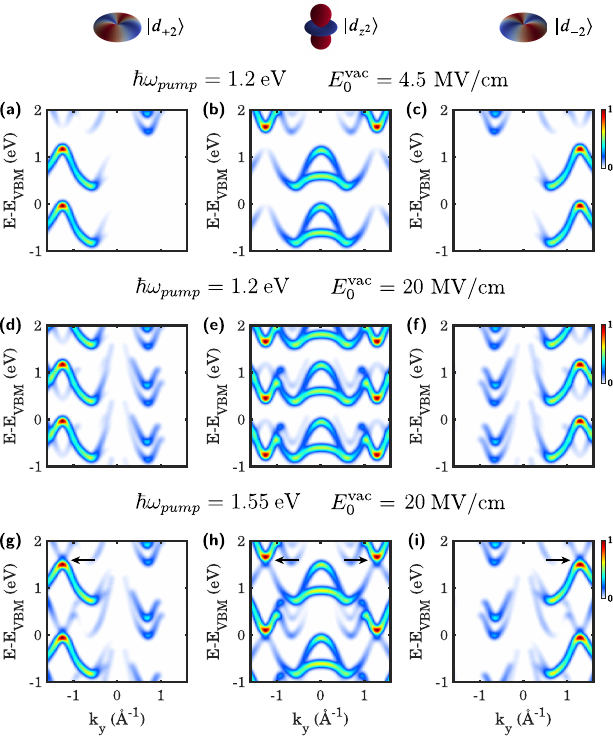}
\caption{\textbf{Orbital-projected Floquet spectral function}. Floquet spectral function projected on \textbf{(a),(d)} and \textbf{(g)}  $d_{+2}$, \textbf{(b),(e)} and \textbf{(h)} $d_{z^2}$, and \textbf{(c),(f)} and \textbf{(i)} $d_{-2}$ orbitals of the top-layer W atoms, using $s-$polarized pump pulse, for three different pumping conditions, i.e. \textbf{(a)-(c)} $\hbar\omega_\mathrm{pump}$ = 1.2~eV, $E_0^\mathrm{vac}$ = 4.5 MV/cm, \textbf{(d)-(f)} $\hbar\omega_\mathrm{pump}$ = 1.2~eV, $E_0^\mathrm{vac}$ = 20 MV/cm, and \textbf{(g)-(i)} $\hbar\omega_\mathrm{pump}$ = 1.55~eV, $E_0^\mathrm{vac}$ = 20 MV/cm. From this orbital-projected Floquet spectral function analysis, we can conclude that within the standard Floquet theory framework, in the off-resonant pumping regime, the orbital character of the sideband is the same as in the valence band, i.e. it does not capture the momentum-dependent hybridization of valence and conduction band measured experimentally and predicted by the time-dependent non-equilibrium Green's function (td-NEGF) calculations. Within this theoretical approach, an appreciable valley-selective Floquet hybridization is only observable for near-resonant strong-field pumping.}
\label{floqorb}
\end{figure}

\begin{figure}
\centering\includegraphics[width=0.55\textwidth]{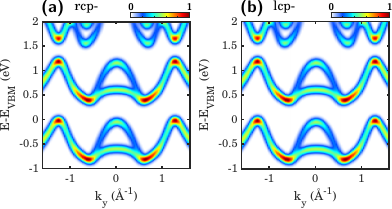}
\caption{\textbf{Floquet spectral function for different helicity of the driving pulse}. Floquet spectral function calculated using \textbf{(a)} right-circularly and \textbf{(b)} left-circularly polarized driving pulses. From this analysis, we can conclude that within the standard Floquet theory framework, Floquet spectral function is independent of the helicity of the driving pulse, i.e. it does not capture the valley-polarized nature of Floquet bands measured experimentally and predicted by the time-dependent non-equilibrium Green's function (td-NEGF) calculations.}
\label{floqchirality}
\end{figure}

\begin{figure}
\centering\includegraphics[width=0.55\textwidth]{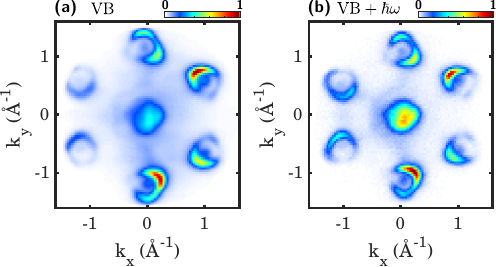}
\caption{\textbf{Valence band and first-order replica constant energy contours}. Constant energy contours obtained by summing the photoemission intensity for all $\theta_{\mathrm{QWP}}^{IR}$ and for $p-$polarized XUV, for the valence band (VB, i.e. $\mathrm{E-E_{VBM}}$ = -0.35~eV) in \textbf{(a)} and for the first-order replica (VB+$\hbar\omega$, i.e. $\mathrm{E-E_{VBM}}$ = 0.85~eV) in \textbf{(b)}.}
\label{CECs}
\end{figure}

\begin{figure}
    \centering
    \includegraphics[width=0.9\textwidth]{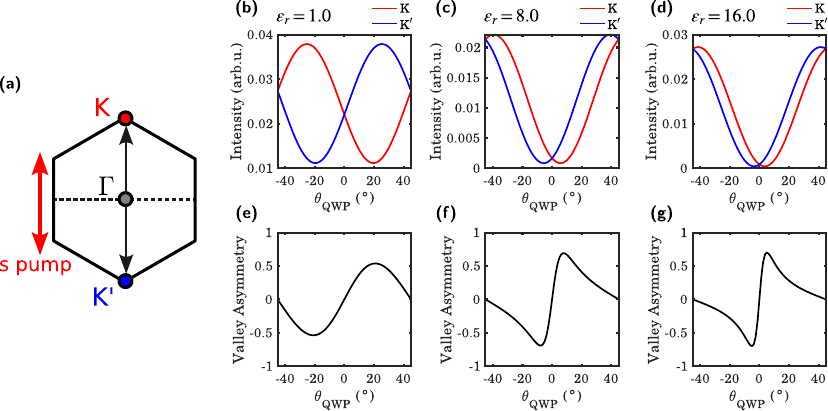}
    \caption{\textbf{Polarization- and Valley-Resolved Volkov Intensity} \textbf{(a)} Scheme of the light-matter interaction geometry, where the light-scattering plane (dashed black line) is aligned along $\Gamma$-M and $s-$polarized pump electric field is aligned along K-$\Gamma$-K$^{\prime}$. \textbf{(b)-(d)} Valley-resolved Volkov intensity and \textbf{(e)-(g)} associated valley asymmetry, as a function of the quarter-wave-plate angle ($\theta_{\mathrm{QWP}}$), for different dielectric constant ($\varepsilon_r$) values. For $\theta_{\mathrm{QWP}}$ = 0$^{\circ}$, pump is $s-$polarized, for $\theta_{\mathrm{QWP}}$ = $\pm$ 45$^{\circ}$, pump is right/left circularly-polarized. These calculations were performed using the simple model inspired by Park~\cite{Park14} described in the Methods.}
    \label{fig:volkov}
\end{figure}

\clearpage

\bibliographystyle{naturemag}
\bibliography{ValleyFloquet.bib}

\end{document}